# Development of a New Gas Puff Imaging Diagnostic on the HL-2A Tokamak

**B. Yuan**[a,b], **M. Xu**[b,*], **Y. Yu**[a,*], **L. Zang**[b], **R. Hong**[c], **C. Chen**[b], **Z. Wang**[b], **L. Nie**[b], **R. Ke**[b], **D. Guo**[b], **Y. Wu**[a,b], **T. Long**[b], **S. Gong**[a,b], **H. Liu**[a,b], **M. Ye**[a], **X. Duan**[b] **and HL-2A team**[b]

[a] *School of Nuclear Science and Technology, USTC, Hefei, China*
[b] *Southwestern Institute of Physics, Chengdu, China*
[c] *Center for Energy Research, University of California, San Diego, United States*
* *Corresponding Author:* `minxu@swip.ac.cn`, `yuyi@ustc.edu.cn`

ABSTRACT: A new gas puff imaging (GPI) diagnostic has been developed on the HL-2A tokamak to study two-dimensional plasma edge turbulence in poloidal vs. radial plane. During a discharge, neutral helium or deuterium gas is puffed at the edge of the plasma through a rectangular multi-capillary nozzle to generate a gas cloud on the observing plane. Then a specially designed telescope and a high-speed camera are used to observe and photograph the emission from the neutral gas cloud. The brightness and contrast in the 2-D poloidal vs. radial frames reveal the structures and movements of the turbulence. The diagnostic was put into the first experiment during the latest campaign and successfully captured blob structures of different shapes and sizes in scrape-off layer (SOL).

KEYWORDS: GPI; turbulence; blob.

# Contents

## 1. Introduction

Edge plasma turbulence is a hot topic in fusion plasma investigations, for it plays an important role in both plasma confinement[1, 2]and plasma-material interaction[3,4]. The edge turbulence determines the transport near the last closed magnetic flux surface (LCFS) or separatrix, and consequently affects the edge temperature profile and density profile. The edge conditions then may affect the core confinement via edge gradients and gradient driven instabilities, such as drift waves[5]. Information of the 2-D structure of plasma edge turbulence can be observed with Langmuir probe array and 2-point correlation techniques[6,7]. However, Langmuir probe cannot penetrate deep enough into plasma, especially when plasma density $n_e$ and temperature $T_e$ are high, e.g. in H-mode discharge.

GPI is a powerful diagnostic that permits a 2D measurement of edge turbulence with high temporal resolution and high spatial resolution[8-12]. In a GPI diagnostic, a controlled neutral gas cloud, of helium or deuterium, is puffed into at the edge of the plasma, and then the visible emission(HeI line at 587.6nm or $D_\alpha$ line at 656.2nm) from the interaction between the neutral particles and plasma is photographed with individual frames. In principle, the emission is modulated instantaneously by both the local plasma density and temperature, which means the structures and movements of the edge turbulence can be inferred from the frames. As edge turbulent structures are usually correlated at long range along magnetic field [13,14], GPI diagnostic is usually used to observe in the poloidal vs. radial plane. Over the past years, GPI diagnostic has been developed in several magnetically confined plasma devices, such as Alcator C-Mod[8], NSTX[8,9], TEXTOR[11] and EAST[10] etc.

A new GPI diagnostic has been developed to study the edge turbulence in the HL-2A tokamak. This paper reports the development of the diagnostic and some initial experimental results.

## 2. Gas puff imaging diagnostic

### 2.1 Principle

In the GPI diagnostic, a controlled neutral gas cloud, of typically helium or deuterium, is puffed at the edge plasma region. Then the visible line emission, the HeI line at 587.6nm or Dα line at 656.2nm, is observed by a telescope and imaged by a high-speed camera or a light-sensitive detector array. Typically, the measurements are done with high spatial resolution and temporal resolution (exposure time of an individual frame, shorter than the autocorrelation time of edge turbulence) to study the structures and movements of plasma turbulence. According to collisional-radiative(CR) models[15, 16], which describe the kinetics of the population of atom levels for typical edge plasma and assume negligible time-dependent effects, the emission depends on the local electron density $n_e$ and temperature $T_e$[17],

$$\epsilon(\text{photons/s } m^3) = n_0 f(n_e, T_e) A,$$

where $n_0$ is the local neutral particle density, A is the radiative decay rate for the observed emission, and $f(n_e, T_e)$ is a nonlinear function of $n_e$ and $T_e$. The decay rate A is much larger than the inverse of the autocorrelation time of the fluctuations (typically larger than 10μs)[8], ensuring that the observed emission corresponds to the local plasma turbulence parameters. As $f(n_e, T_e)$ indicates, the emission from the plasma is modulated by both the local electron density $n_e$ and temperature $T_e$ fluctuations and the actual dependence on $n_e$ and $T_e$ varies with different plasma conditions as well as different injected neutral gas. The results of GPI diagnostic and Langmuir probes experiments on the TEXTOR tokamak prove that in the edge region the emission depends more on local electron density than temperature, especially at high frequency band[11].

### 2.2 Diagnostic Setup

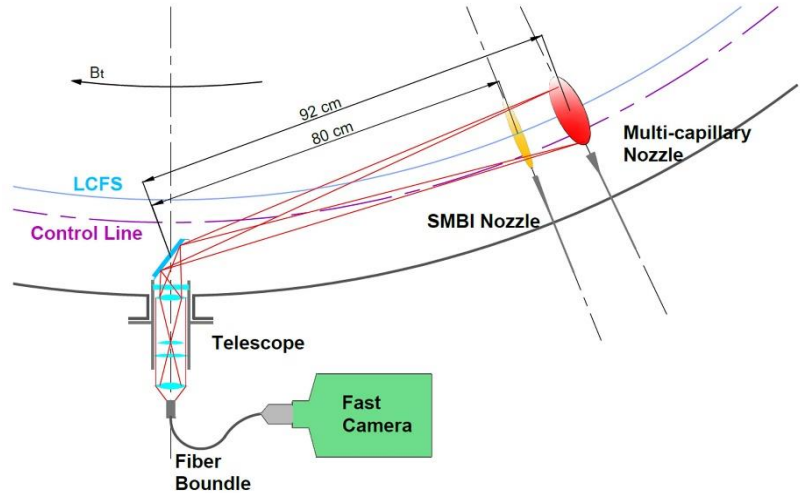

**Figure 1.** The diagrammatic sketch (top view) of the GPI diagnostic on the HL-2A tokamak.

A GPI diagnostic with high speed camera has been developed on the HL-2A tokamak[18], a conventional tokamak (minor radius a = 40cm, major radius R = 165cm) that operates in limiter mode or diverter mode. In experiments, typical plasma current is $I_P = (150 − 450)$kA, toroidal magnetic field $B_T = (1.35 − 2.7)$T and line-average electron density $\bar{n}_{e0} = (1 − 8) \times 10^{19} m^{-3}$. Figure 1 shows the diagrammatic sketch of the GPI diagnostic on the HL-2A tokamak. A neutral gas cloud is puffed at the edge of the plasma through a nozzle installed on the vacuum chamber wall. The emission from the gas cloud is observed by the telescope and guided to the fast camera.

Two rectangular multi-capillaries nozzles were installed next to the SMBI nozzle, 92cm away from the telescope lens, 5cm above the mid-plane, and the front surfaces of the nozzles are at the minor radius of a = 46cm. The two nozzles are different in size and capillary distribution to produce gas cloud in different shapes and profiles. One is 2cm x 6cm with $5 \times 20$ capillaries on it, and each capillary is 3mm away from the others in two directions. And the other one is 2cm × 14cm with $2 \times 27$ capillaries on it, and the distances in two directions are 10mm and 5mm respectively. All the capillaries on the two nozzles have the diameter of $\phi = 0.5$mm, the length of $L = 10$mm, and a large aspect ratio of $A = L/\phi = 20$, which contributes to reducing gas divergence. The two nozzles are connected to a PEV-1 piezoelectric valve with two 2m-long Z-shape stainless steel tubes. The piezoelectric valve is driven by 0-100V DC and responses in about 2ms typically. A 2-litres gas reservoir is used as an intermediate chamber to store the working gas under a preset pressure range from 1.0 to 2.0atm before each puffing pulse.

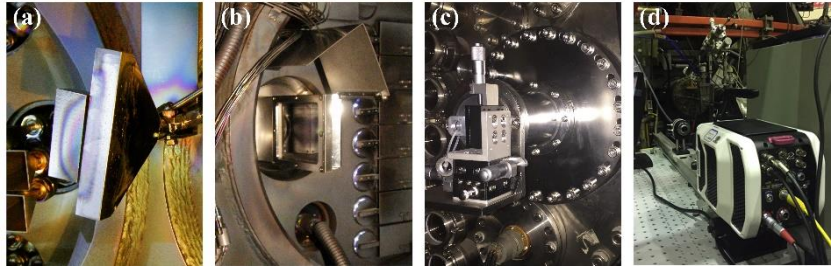

**Figure 2.** (a) Two rectangular multi-capillaries nozzles are installed on the vacuum chamber wall with a 6° clockwise deflection to be perpendicular to the local magnetic field. (b) The telescope penetrates into the vacuum chamber (the shutter is open). (c)A 3-axis stage is at the outside end of the telescope to hold the imaging optical fiber bundle. (d)The Phantom V2011 high-speed camera.

The optic system, as figure 1 shows, consists of a front telescope, an optical fiber bundle, relay lenses, narrow band interference filter, and a high speed camera.

The front telescope is a custom-made one with the focal length of 30mm, and the numerical aperture of F/1.01, and is installed on a mid-plane port in the ONO sector, next to the HCN interferometer diagnostic. Inside the vacuum chamber, a reflecting mirror with silver coating on the surface is at the head of the telescope, and a rotatable protecting shutter is used to keep the first mirror from being coated during glow discharge cleanings or other irrelevant experiments. The imaging lens module is located on the outside of the flange, and it can be removed without opening the vacuum chamber. At the end of the lens module is a 3-axis stage, which is used to hold the imaging optical fiber bundle.

The optical fiber bundle has 400 pixels × 400 pixels on its 4mm × 4mm imaging plane, correspondingly covering a 160mm × 160mm square at the gas cloud plane (140mm × 140mm at the SMBI beam plane). And the imaging area can be moved along radial direction by moving the optical fiber bundle head on the 3-axis stage, covering from a=25cm( ρ~0.625 ) to

a=45cm($\rho \sim 1.125$), to follow the separatrix as it shifts from limiter discharge to diverter discharge. Narrow band (FWHM=1.5nm) interference filters with central wavelengths of λ=656.2nm (Da line) or λ=587.6nm (He I line) are utilized to eliminate the influence from the light of nearby wavelengths.

At the end of the optic path is a Phantom V2011 high-speed camera to image the emission with 128 pixels × 128 pixels, and thus the spatial resolution at the observing plane is 1.25mm × 1.25mm. When it runs with 128 pixels × 128 pixels, the sample rate can reach up to 420,000 fps, and the minimum exposure time for a single frame is 1μs(the actual exposure time may be several microseconds, depending on the emission intensity).

## 3. First Experimental Results

During the latest experiment campaign of HL-2A, the newly developed GPI diagnostic was put into experiment and got some initial results of several shots. Because these shots were carried out under ohmic discharges without any auxiliary heating, the results only proved the diagnostic's ability to capture a series of physical phenomena at edge. And in this paper, the discussion will not go deep into physics issues and corresponding mechanisms.

During an typical ohmic diverter deuterium discharge No.31886 with $B_t = 1.32$ T, $I_P = 144$ kA and $\bar{n}_{e0} = 1.2 \times 10^{19}$ m$^{-3}$, a helium gas puffing pulse for 10ms, pressure of $1.95 \times 10^5$ Pa, was puffed into the plasma through the 2cm ×14cm nozzle at 300ms, and the neutral gas flux was estimated to be $2.5 \times 10^{20}$ molecule · s$^{-1}$ in average. After the puffing pulse, the observing area went bright about 10ms later. Figure 3 (a) shows a single raw frame at 315ms with 9μs exposure time, and the poloidal ion diamagnetic drift direction in the scrape-off layer (SOL) is shown by the arrow on the right. Figure 3 (b) is the average emission intensity of 500 frames from 315ms to 320ms, which indicates that the emission is mainly located in an 8cm-wide band (marked in green and yellow) on the both sides of the separatrix (marked as the white dashed line).

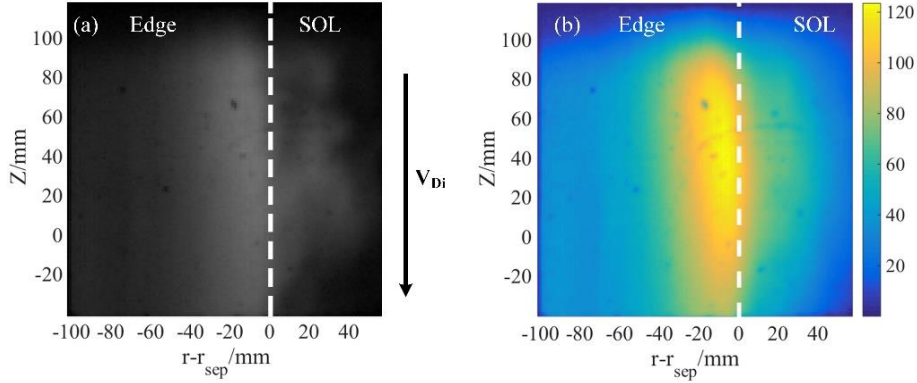

**Figure 3.** (a)A single raw frame from 315ms with 9μs exposure time during a typical ohmic diverter deuterium discharge No.31886 ($B_t = 1.32$T, $I_P = 144$kA and $\bar{n}_{e0} = 1.2 \times 10^{19}$m$^{-3}$). The observing area is a 16cm × 16cm square in the poloidal vs radial plane, and the separatrix is marked as a white dashed line. (b)The average emission of 500 continuous frame from 315ms to 316ms.

Figure 4 shows 3 continuous frames from 315.16ms to 315.19ms with an interval time of 10μs and an exposure time of 9μs. To suppress thermal noise and heighten image contrast, the frames were processed with a 2-D median-filter (smooth the signal in 3pixels × 3pixels) and normalized by the average intensity. The bright structures in the SOL region represent blobs of different sizes and shapes, and all the blobs are moving downward in poloidal direction (indicated by yellow dashed line and green dashed line respectively) and outward in radial direction. Both two marked structures change shapes as they move outward and downward. And the small one (marked by green dashed line) moves faster in poloidal direction than the large one.

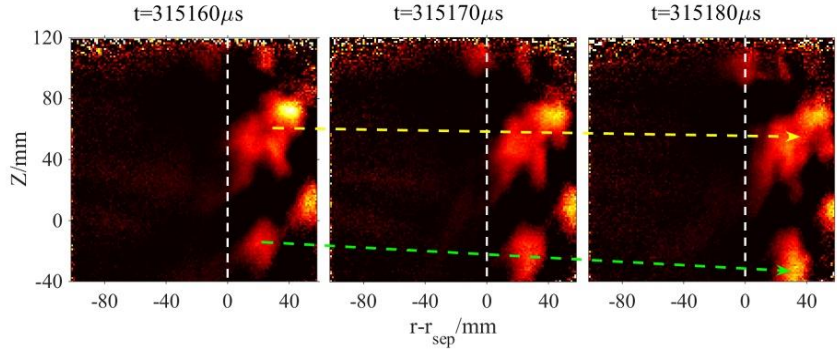

**Figure 4.** A series of 3 continuous frames taken from 315.16ms to 315.19ms during the discharge No.31886. A yellow dashed line and a green dashed line indicate the poloidal movements of two different size blob structures respectively.

Though the emission band extends several centimeters inside the separatrix, structures inside the separatrix are hard to recognize. The reason might be that the GPI diagnostic was running with a comparative low framing rate of 100kfps, and thus the corresponding Nyquist frequency was 50 kHz. While plasma turbulence inside the separatrix extends to higher frequency (e.g. 70 to 100 kHz), so there the GPI diagnostic only recorded a smoothed signal, which seems like a misty emission band in Figure 3(a).

## 4. Summary

The newly developed GPI diagnostic on the HL-2A tokamak has been put into the first experiment in the latest campaign. During a typical ohmic diverter deuterium discharge No.31886, the diagnostic successfully captured several blob structures of different sizes in the poloidal vs radial plane with a spatial resolution of 1.25mm × 1.25mm and a temporal resolution of 10μs. The sizes and movements of the structures can be inferred from a series of frames. But during this campaign, the diagnostic was running at a framing rate of 100kfps, which is correspondingly low for turbulence measurement inside the separatrix. To minimized the perturbation to the local plasma, the gas puffing pulse was controlled at a low flux of about $2.5 \times 10^{20}$ molecule $\cdot$ s$^{-1}$.

So for the next experimental campaign, the framing rate will be improved to more than 200kfps to capture faster phenomena. Meanwhile the experimental results will be compared with those from other diagnostics, e.g. Beam Emission Spectroscopy (BES), Langmuir Probes and Doppler Reflectometer, to calibrate the measurements and to identify both the local and non-local perturbations introduced by the neutral gas puff. And the GPI diagnostic is expected to measure plasmas with auxiliary heating as well as H-mode discharges. Besides the experimental work, computational simulations, e.g. BOUT++ code, will contribute to better interpretation of the GPI data by producing deposition and emission profiles of neutral gas.

**5. Acknowledgements**

We are grateful for the contribution of the HL-2A staff. The development of the GPI diagnostic was supported by the National Natural Science Foundation of China under Grant No.11575055, No.11505052 and the Chinese National Fusion Project for ITER under Grant No.2013GB107001, 2013GB111005.